# Design Considerations of a Coordinative Demand Charge Mitigation Strategy

Rongxing Hu, Kai Ye, Hyeonjin Kim, Hanpyo Lee, Ning Lu
North Carolina State University
Raleigh, NC 27606, USA
{rhu5, kye3, hkim66, hlee39, nlu2}@ncsu.edu

Di Wu
Pacific Northwest National Laboratory
Richland, WA 99352, USA
di.wu@pnnl.gov

PJ Rehm
ElectriCities of North Carolina Inc
Raleigh, NC 27604, USA
prehm@electricities.org

*Abstract*—This paper presents a coordinative demand charge mitigation (DCM) strategy for reducing electricity consumption during system peak periods. Available DCM resources include batteries, diesel generators, controllable appliance loads, and conservation voltage reduction. All resources are directly controlled by load serving entities. A mixed integer linear programming-based energy management algorithm is developed to optimally coordinate DCM resources considering the load payback effect. To better capture system peak periods, two different kinds of load forecast are used: the day-ahead load forecast and the peak-hour probability forecast. Five DCM strategies are compared for reconciling the discrepancy between the two forecasting results. The DCM strategies are tested using actual utility data. Simulation results show that the proposed algorithm can effectively mitigate the demand charge while preventing the system peak from being shifted to the payback hours. We also identify the diminishing return effect, which can help load serving entities optimize the size of their DCM resources.

## I. Introduction

Many load serving entities (LSEs), especially municipal utilities and electric cooperatives, are billed a demand charge (or capacity charge) and energy charge when purchasing electricity from the wholesale electricity market or from a large utility. Demand charges are mainly designed to recover the investment in electricity generation and transportation infrastructure, and to encourage the users to reduce the coincident peak demand. As a result, electricity consumed during the system's monthly peak hour can be charged at a very high demand charge rate (e.g., 20 $/kW-month). For many LSEs, demand charge payment can account for over 50% of their monthly wholesale power bill. This incentivizes LSEs to deploy demand charge mitigation (DCM) strategies for load reduction during forecasted system peak load periods. Commonly used DCM resources include battery energy storage systems (BESS), diesel generators (DG), thermostatically controlled loads (TCL), and conservation voltage reduction (CVR).

In the past, because the monthly peak usually occurs inside the forecasted system peak load periods, DCM is a trivial optimization problem to solve. Utilities will execute load reduction only in hours with forecasted load peak. However, in the past two years, when more and more LSEs started to deploy DCM programs, the system load shape has been flattened. System peaks can occur in a wider range of hours or be shifted to previously non-peak hours. Meanwhile, high penetration of variable renewable generation resources, especially the integration of distributed roof-top PV systems, can also shift the system load peak to unexpected hours (e.g., a sudden drop of PV outputs due to cloud covers).

In 2022, many LSEs in North Carolina found that some monthly peaks occurred outside their forecasted peak periods. However, prolonged operation of DCM resources requires additional cost, and many DCM resources are constrained by power and energy operation limits, making the development of optimization-based energy management algorithms necessary.

Many existing DCM methods use predefined demand thresholds to trigger DCM. In [1], Moslemi *et al.* propose to manage behind-the-meter storage for DCM based on achievable monthly demand thresholds obtained using a data-driven approach. In [2], Ng *et al.* present to update the demand threshold using the day-ahead load profile if long-term historical data are not available. In [3], Pflaum *et al.* use a linear programming model to control batteries and generators for DCM using load profiles from the previous year as inputs. The threshold based method is also used to control electric vehicle charging [4] or co-optimize BESS and TCLs [5] for DCM. Recently, in [6], Wu *et al.* develop a two-step control and assessment framework for controlling BESSs for DCM based on peak-day and peak-hour probabilities.

We identify two drawbacks in the existing approaches. First, forecast reconciliation between the day-ahead load forecast and peak probability forecast is not considered. Second, the payback period [7,8] when using TCL loads for DCM is not considered. Thus, in this paper, we present a coordinative DCM strategy for reducing electricity consumption during system peak periods to hedge demand charges. Our innovations are threefold. *First*, we proposed a mixed integer linear programming (MILP) based DCM management algorithm to coordinate demand reduction considering load payback effects. *Second*, we proposed five DCM strategies to reconcile the forecasting discrepancy between day-ahead load forecast and peak day/hour probability forecast. *Third*, we identify the diminishing return effect that can help LSEs size their DCM programs. The proposed method is tested using actual utility operational data. Simulation results show that the proposed DCM algorithm can effectively mitigate the demand charge while avoiding shifting system peaks to unexpected hours due to the payback of TCLs.

## II. METHODOLOGY

In this section, we present the workflow of the proposed algorithm, DCM resource models, and DCM strategies.

### A. An Overview of the Coordinative DCM Algorithm

As shown in Fig. 1, for the $i^{th}$ day in a month, the forecasted peak load ($P_i^{peak}$) is first compared with the historical peak load ($P_{his}^{peak}$) of the month. Note that the peak demand of the LSEs is the maximum hourly load in a month. Let $K_{error}$ be an error margin (e.g., 10%). If $P_i^{peak}(1 + K_{error}) > P_{his}^{peak}$, the peak day probability ($p_i^{peak}$) will be compared with the predefined threshold ($p_{th}^{peak}$). If $p_i^{peak} \geq p_{th}^{peak}$, the coordinative DCM will be executed. We use two load forecasts for targeted hour selection: the peak hour forecast introduced in [6] and the day-ahead system load forecast.

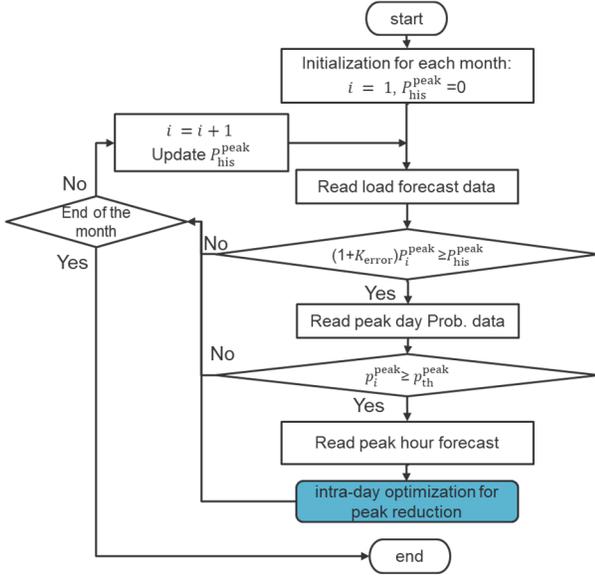

Fig. 1. Flowchart of the coordinative DCM algorithm.

### B. Demand Mitigation Resources

The DCM resources include BESS, DG, TCL, and CVR. Note that all resources are directly controlled by the LSE.

*1) BESS Model:* We assume that the BESS is fully charged (SOC=100%) and ready for DCM at the beginning of the day and the BESS will be solely used for DCM. Thus, the BESS will only discharge during the DCM periods and is modeled by

$$E_t = E_{t-1} - \frac{P_t^{bat}}{\eta}\Delta t, \forall t \in \mathcal{T} \quad (1)$$

$$E_{min} \leq E_t \leq E_{max} \quad (2)$$

$$0 \leq P_t^{bat} \leq P_{max}^{bat} \quad (3)$$

where $E_t$ is the energy storage at the end of time $t$; $E_{min}$ and $E_{max}$ are the minimum and maximum energy limits, respectively; $P_t^{bat}$ is the power discharged to the grid; $P_{max}^{bat}$ is the maximum discharge power; $\eta$ is the discharging efficiency; $\mathcal{T}$ (dimension is $N_{hour} \times 1$) denotes the set of targeted demand reduction hours, where $N_{hour}$ is the number of the targeted demand reduction hours.

*2) DG Model.* The cost of DG for DCM is calculated as

$$C^{dg} = \gamma^{fuel}\sum_{t\in\mathcal{T}}P_t^{dg}\Delta t - \alpha^{eng}\sum_{t\in\mathcal{T}}P_t^{dg}\Delta t \quad (4)$$

$$0 \leq P_t^{dg} \leq P_{max}^{dg} \quad (5)$$

where $C^{dg}$ is the DG's demand reduction cost; $P_t^{dg}$ is the DG power output at time $t$; $P_{max}^{dg}$ is the maximum DG power; $\gamma^{fuel}$ is the fuel cost, 0.245$/kWh; $\alpha^{eng}$ is the energy purchase price, 0.03 $/kWh in this paper.

*3) CVR Model.* The effective CVR deployment is only for a limited period [9]. We assume that there is 1 aggregated CVR resource in the system, the CVR can be executed for a period of 3-hours or less continuously. After one execution period, CVR requires at least 1 hour to regain its effectiveness. Define $\boldsymbol{U}^{CVR}$ as the selection matrix, where $u_j^{CVR} = 1$ means the option $j$ is selected and $u_j^{CVR} = 0$ means not selected. Define $\boldsymbol{M}^{CVR}$ (dimension is $N_{hour} \times N_{CVR}$) as the option matrix. In total, we have $N_{CVR}$ feasible CVR execution options. Each column in $\boldsymbol{M}^{CVR}$ denotes the on/off status of the CVR program at each of the $N_{hour}$ targeted demand reduction hours. Let $u_t^{cvr*}$ denote the CVR status at hour $t$, then the final selected CVR operation option is expressed as

$$\left[u_1^{CVR*}, u_2^{CVR*} \dots u_{N_{hour}}^{CVR*}\right]^T = \boldsymbol{M}^{CVR}\left[u_1^{CVR}, u_2^{CVR} \dots u_{N_{CVR}}^{CVR}\right]^T \quad (6)$$

$$\sum_{j=1}^{N_{CVR}} u_j^{CVR} = 1 \quad (7)$$

As shown in Fig. 2, the values in $\boldsymbol{M}^{CVR}$ for a day with 4 possible targeted load peak hours and 16 operation options. If the selection matrix selects option 4, the targeted hours will be hours 15:00, 17:00, and 19:00.

| | Targeted hours | \multicolumn{16}{c|}{options} |
|---|---|---|---|---|---|---|---|---|---|---|---|---|---|---|---|---|---|
| | | 1 | 2 | 3 | 4 | 5 | 6 | 7 | 8 | 9 | 10 | 11 | 12 | 13 | 14 | 15 | 16 |
| $\boldsymbol{M}^{CVR}$ | 15:00 | 1 | 1 | 1 | 1 | 0 | 1 | 1 | 1 | 0 | 0 | 1 | 0 | 0 | 0 | 0 | 0 |
| | 16:00 | 1 | 1 | 1 | 0 | 1 | 1 | 0 | 0 | 1 | 0 | 1 | 0 | 1 | 0 | 0 | 0 |
| | 17:00 | 1 | 1 | 0 | 1 | 1 | 0 | 1 | 0 | 1 | 1 | 0 | 0 | 0 | 1 | 0 | 0 |
| | 19:00 | 1 | 0 | 1 | 1 | 1 | 0 | 0 | 1 | 0 | 1 | 1 | 0 | 0 | 0 | 1 | 0 |
| $[u_1^{CVR}, u_2^{CVR} \dots u_{N_{CVR}}^{CVR}]$ | | 0 | 0 | 0 | 1 | 0 | 0 | 0 | 0 | 0 | 0 | 0 | 0 | 0 | 0 | 0 | 0 |
| $[u_1^{CVR*}, u_2^{CVR*} \dots u_{N_{hour}}^{CVR*}]$ | | 1 | 0 | 1 | 1 | | | | | | | | | | | | |

Fig. 2. An example of the option matrix, the selection matrix, and the selected CVR operation option.

The CVR load reduction at time $t$, $\Delta P_t^{cvr}$, is calculated as

$$\Delta P_t^{CVR} = u_t^{CVR*}k_1^{CVR}k_2^{CVR}P_t^{load} \quad (8)$$

where $k_1^{CVR}$ is the percentage of the total load at time $t$ ($P_t^{load}$) that can be used for CVR; $k_2^{CVR}$ is the CVR load reduction factor.

*4) TCL Model.* In this paper, we consider two TCL-based demand response (DR) groups. The first group of HVACs follows a cycling rule of 10-minutes off in 30-minutes and the second group of HVACs is turned off for the entire 30-minutes. There is no time limit for the first DR group. For the second group, the DR can only be executed for no more than 2 consecutive hours. Note that the TCL-based DR will be followed by a payback period as explained in [7,8]. As shown in Fig. 3, if not well managed, the payback period may create a new system load peak. Thus, it is necessary to account for payback when using DR for demand reduction.

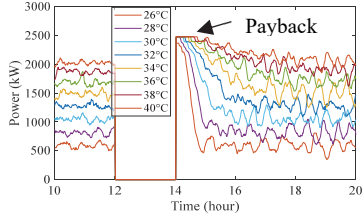

Fig. 3. Example of HVAC load payback effects.

To estimate the payback effect of each DR group, we model the HVAC load using the modeling method from [10] based on data acquired from HVAC load disaggregation algorithms [11]. First, the normal HVAC load consumption is modeled using the forecasted ambient temperature profile as the input. Next, the load profiles of the two DR groups are modeled for all deployment options. Finally, the demand reduction is calculated by comparing the normal HVAC load profile with the load profile for each DR deployment option.

Denote $M_s^{Ptcl}$ and $M_s^{Htcl}$ as the *DR power reduction* option matrix and the *DR deployment* option matrix, respectively. The dimension of the matrices is $N_{hour} \times N_s^{tcl}$ containing $N_s^{tcl}$ options for the $N_{hour}$ targeted hours. Then, the TCL option selection process is formulated as

$$[\Delta P_{s,1}^{tcl}, \Delta P_{s,2}^{tcl} \ldots \Delta P_{s,N_{hour}}^{tcl}]^T = M_s^{Ptcl}[u_{s,1}^{tcl}, u_{s,2}^{tcl} \ldots u_{s,N_s^{tcl}}^{tcl}]^T \quad (9)$$

$$\sum_{z=1}^{N_s^{tcl}} U_{s,z}^{tcl} = 1 \quad (10)$$

$$\Delta P_t^{tcl} = \sum_{s=1}^{N_s} \pi_s^{tcl} \Delta P_{s,t}^{tcl} \quad (11)$$

$$[u_{s,1}^{tcl*}, u_{s,2}^{tcl*} \ldots u_{s,N_{hour}}^{tcl*}]^T = M_s^{Htcl}[u_{s,1}^{tcl}, u_{s,2}^{tcl} \ldots u_{s,N_s^{tcl}}^{tcl}]^T \quad (12)$$

$$C^{tcl} = \sum_{t \in \mathcal{T}} \sum_{s=1}^{N_s} u_{s,t}^{tcl*} \quad (13)$$

where $s$ denotes the DR group, $N_s$ is the total number of DR groups, $N_s^{tcl}$ is the number of DR deployment options for the $s^{th}$ DR group; $U_{s,z}^{tcl}$ represents the deployment status of the $z^{th}$ option for the DR group $s$; $u_{s,t}^{tcl*}$ denotes the selected DR status in hour $t$; $\Delta P_{s,t}^{tcl}$ represents the demand reduction; $\pi_s^{tcl}$ is the scale factor which scales up the number of the selected typical HVACs in DR simulation to that of the field; $C^{tcl}$ is the total number of hours when DR is deployed.

*C. Intra-day Optimization*

Two different types of load forecasts are used for identifying possible peak-load hours: the 24-hour ahead load forecast and the machine-learning based peak-hour likelihood forecast [6]. As shown in Fig. 4, in a day, only a few hours are possible daily peak hours, thereby possible monthly peak hours. To dispatch limited DCM resources, we design and compare five strategies (S1-S5) based on how an LSE selects the DCM hours. Note that the strategy for DCM hour selection reflects how much risk an LSE is willing to take for missing an actual monthly peak.

*1) S1 (Prob-TopX).* Select X hours ($\mathcal{T}_{top}^{prob}$) based on the probability of being the peak hour. The objective function is:

$$max. \ f_1 = \beta \sum_{t \in \mathcal{T}_{top}^{prob}} \mu_t \Delta P_t - C^{od} \quad (14)$$

$$\mu_t = \frac{p_t}{\sum_{t \in \mathcal{T}_{top}^{prob}} p_t} \quad (15)$$

$$\Delta P_t = P_t^{bat} + P_t^{dg} + \Delta P_t^{CVR} + \Delta P_t^{tcl}, \forall t \in \mathcal{T}_{top}^{prob} \quad (16)$$

$$C^{od} = C^{dg} + C^{tcl} \quad (17)$$

where $\mu_t$ is the normalized factor based on the peak hour probability and $C^{od}$ denotes the penalty factor which considers the DG fuel cost and the total number of HVAC DR hours for taking the customer comfort needs into account.

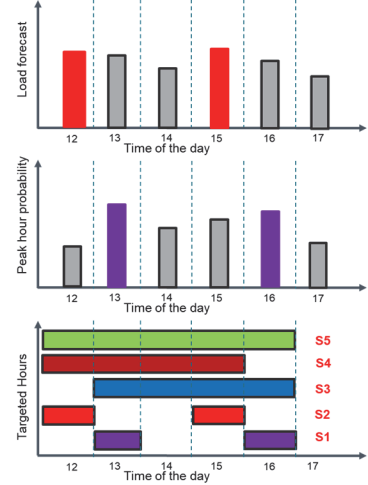

Fig. 4. (a) Peak hours forecasted by the day-ahead load forecast, (b) Peak hours forecasted by the machine-learning based peak-hour likelihood forecast, and (c) Target hours selected by the given DCM strategies.

*2) S2 (DALF-TopX).* Select X hours ($\mathcal{T}_{top}^{pred}$) with the highest load based on the day-ahead load forecast. The objective function is:

$$max. \ f_2 = \beta \Delta P^{peak} - C^{od} \quad (18)$$

$$\Delta P^{peak} = P_t^{pred} - P^{peak}, t = t_{peak}^{pred} \quad (19)$$

$$P^{peak} \geq P_t, \forall t \in \mathcal{T}_{top}^{pred} \quad (20)$$

$$P_t = P_t^{pred} - P_t^{bat} - P_t^{dg} - \Delta P_t^{CVR} - \Delta P_t^{tcl}, \forall t \in \mathcal{T}_{top}^{pred} \quad (21)$$

where $\Delta P^{peak}$ is the expected peak load reduction in the day, $P_t^{pred}$ denotes the load forecast, $P^{peak}$ is the expected peak load after peak reduction, $t_{peak}^{pred}$ is the hour when the forecasted peak load occurs. Equation (19) calculates the expected peak load reduction, (20) obtains the expected peak load after peak reduction, (21) denotes the power balance

*3) S3 (Prob-Horizon).* The peak reduction is executed for a continuous time horizon ($\mathcal{T}_{horiz}^{prob}$) that includes hours between the X targeted hours. Note that the formulation is similar to that of S2, but there can be more hours in $\mathcal{T}_{horiz}^{prob}$.

*4) S4 (DALF-Horizon).* The peak reduction is executed for a horizon ($\mathcal{T}_{horiz}^{pred}$) that includes hours between the X targeted hours. The formulation is similar to S2, but there can be more hours in $\mathcal{T}_{horiz}^{pred}$.

*5) S5 (CombinedHorizon).* The peak reduction is executed for a horizon ($\mathcal{T}_{horiz}^{comb}$) that includes all hours between top-X peak probability and top-X load prediction. The formulation is similar to S2, but there may be more hours in $\mathcal{T}_{horiz}^{comb}$.

For horizon-based strategies, i.e., S3, S4, S5, an extra hour can be added to the end of the DR hour to handle the impact of the DR payback. Also, in this paper, we set $X = 2$ so that only the top2 peak hours (probability or load forecast) are considered.

## III. CASE STUDY

The performance of the proposed approach is demonstrated using data and rate structure in North Carolina. North Carolina Eastern Municipal Power Agency (NCEMPA) provides wholesale power for 32 cities and towns in eastern North Carolina. To provide this service, NCEMPA has entered into a long-term Full Requirements Power Purchase Agreement (FRPPA) with Duke Energy Progress (DEP). Power purchases under the FRPPA are divided into energy and capacity purchases. Capacity is purchased on a monthly basis, and the capacity purchase is determined by the total power delivered to all 32 Members during DEP's peak hour of consumption each month. The resulting capacity charge for NCEMPA equals the product of the Member's demand during DEP's peak monthly hour and the capacity rate formulated under the FRPPA, which can be greater than \$20/kW-month. Thus, through the FRPPA, NCEMPA can reduce its peak consumption during the monthly peak hour through CVR, DR programs of HVACs and water heaters, DGs, and utility-scale batteries.

In this paper, to demonstrate the algorithm performance in a realistic setting, we use the FRPPA pricing mechanism to set up our use cases. Thus, the monthly demand is calculated as the total amount of electricity consumption (hourly average power) during the system's monthly peak hour. The demand charge rate is set to be \$20/kW-month. The main DCM resources considered include a DG (40-MW), a BESS (2 hours @ 10-MW), 2 DR groups (each has 30,000 HVAC units), and CVR, the setting of which is similar to that of the actual NCEMPA DCM program. In this study, $k_1^{CVR} = 40\%$, $k_2^{CVR} = 1\%$. The DR program is deployed in four summer months (i.e., June, July, August, and September). The load forecasts and actual load profiles of DEP are at 60-minute resolution and provided by the Energy Information Administration (EIA). The forecast of the peak day probability and the peak hour probability in [6] are used. The weather data collected at the Raleigh weather station by the National Oceanic and Atmospheric Administration is used for the HVAC models for predicting the payback effects.

### A. Performance Comparison of the Five DCM Strategies

BESS and DG are two typical DCM resources. The use of BESS for DCM is limited by its energy storage limit but there is no limitation in using DG since the demand reduction period is usually shorter than the DG fuel storage capacity. First, we tested the proposed strategies for the 2-hour BESS only and the DG only cases when their power rating increases from 100 MW to 500 MW at 100 MW increment for six 6 years.

*1) Annual DCM savings.* We run the five DCM algorithms on historical load data and compare their performance by calculating the normalized savings (divide the annual DCM savings of a case by the largest DCM savings of all five strategies). In Figs. 5 and 6, we use color boxes to represent the normalized savings. We compare two base cases: using BESS only and using DG only. In Fig. 5, S1 (*Prob-Top2*) consistently shows good performance over other strategies across the six testing years (except 2001) and when the BESS capacity is increased from 100 MW to 500 MW. This shows that using the peak hour probability forecast (S1) as inputs captures the monthly peak hours better than using the day-ahead load forecast (S2 and S4). S5 (*CombinedHorizon*) is the second best strategy because it accounts for both the load forecast and the peak demand probability forecast. Moreover, S5 performs better at higher BESS power ratings. The DG results (Fig. 6) also show that S1 and S5 are the top-2 strategies. However, compared to BESS, DG has lower savings due to additional fuel cost and the disadvantage exacerbates when using a larger DG.

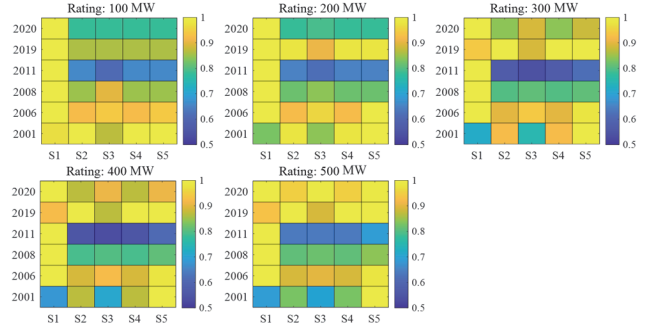

Fig. 5. DCM Performance of BESS.

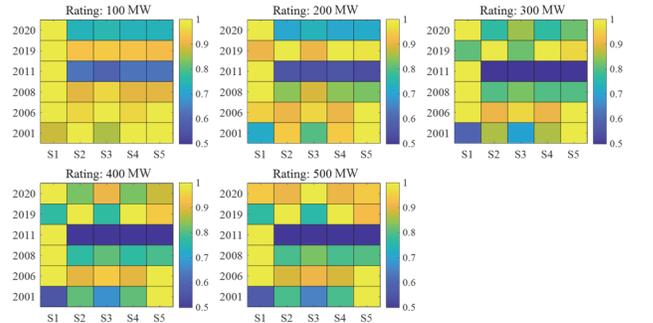

Fig. 6. DCM Performance of DG.

*2) Monthly performance.* As shown in Fig. 7, the load profile varies each month. The summer months have one peak period, but the winter months may have two peak periods. To investigate how load shapes affect the DCM performance, we further compare the performance of each strategy on a monthly basis. In Table I, we first highlight the strategy with the best performance. Then, if the second best strategy achieves very similar performance as that of the best strategy, we also mark it in the table. Again, we find that S1 is the best strategy in most months and S5 is the second best one. However, S5 shows the best performance for the 4 summer months when the two DR control groups are deployed. The results indicate that S5 can help reconcile different demand reduction strategies better by considering the payback effect in summer months.

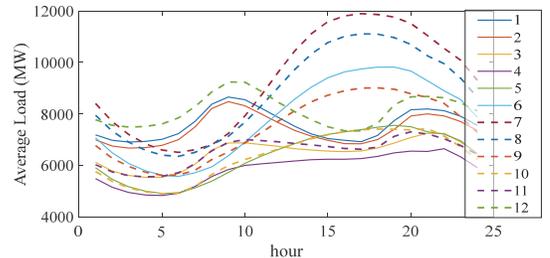

Fig. 7. Load profile in each month in 2020.

## B. Coordinative Demand Mitigation

Based on the performance comparison of the proposed five DCM strategies, S5 is preferred for summer months and S1 for other months in year 2020. In this section, we first evaluate the DCM performance when using the existing utility DCM resource combinations. The annual cost reduction is presented in Fig. 8(a). Because DG and CVR have the largest deployment capacity, their contributions are the largest. Although the two HVAC load groups are deployed only in the summer months, they still achieve significant demand reduction.

Next, by doubling the number of HVACs, we assess whether or not payback can shift the system peak to another hour. As shown in Fig. 8(b), after the targeted DR period (3 hours) ends at 19:00 p.m., a payback peak is observed at 20:00 p.m. To avoid generating a new peak at 20:00 p.m., although the first DR group (with 10 min off in 30 min) does not have limitations in execution duration, the algorithm only deploys it for 2 hours.

TABLE I
MONTHLY PERFORMANCE FOR BESS

|    | Month in a year |   |   |   |   |   |   |   |   |    |    |    |
|----|---|---|---|---|---|---|---|---|---|----|----|----|
|    | 1 | 2 | 3 | 4 | 5 | 6 | 7 | 8 | 9 | 10 | 11 | 12 |
| S1 | T | T | T | T |   | T |   | T |   | T  | T  | T  |
| S2 |   |   |   |   |   |   |   |   |   |    |    |    |
| S3 |   |   |   |   |   |   |   |   |   |    |    |    |
| S4 |   |   |   |   |   | T |   | T |   |    |    |    |
| S5 |   |   |   |   |   | T | **T** | **T** | **T** | T  | T  |    |

## C. Sensitivity Analysis

Using the proposed algorithm, we conduct extensive cost-benefit studies on the sizing of the DCM program. Note that because the BESS degradation is highly correlated to its charge cycles, we use the DCM reduction per battery cycle to estimate the marginal benefit of using BESS for DCM.

As shown in Fig. 9, the yearly DCM benefit decreases when the power rating increases, showing that the incremental benefit is diminishing. For the first 400 MW (about 3% of the peak load), the DCM benefit is prominent. After 800 MW (about 6% of the peak load), the marginal benefit diminishes quickly when adding more DGs and BESSs for DCM.

The reasons for the benefit diminishing effect are two-fold. *First*, the system load peaks have been flattened so that more DCM resources are needed to further flatten the peak load. In addition, loads may need to be shifted further to valley hours instead of shoulder hours. *Second*, although increasing the size of DCM resources can lead to more load reductions in the targeted DCM hours, it also increases the likelihood for the system load peak to be shifted to previously non-peak hours. This makes load reduction in one or two forecasted peak hours no longer effective.

## IV. CONCLUSION

In this paper, we present an energy management method for LSEs to reduce peak demand charge using multiple distributed energy resources, including BESS, DG, TCL, and CVR. The proposed algorithm considers the coordination among those resources while accounting for the payback effect of the TCL DR program. By reconciliation of the day-ahead load forecast and load peak probability forecast, the chance for capturing the system peak load periods is increased. Our cost-benefit study results show that the benefit diminishes when increasing the size of DCM resources. We further demonstrate that there exists a sweet spot when deploying DCM programs, which results in the optimum balance of costs and benefits.

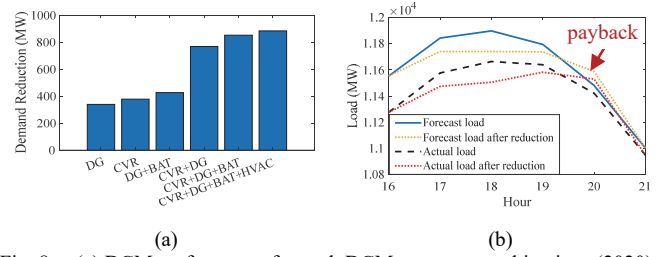

Fig. 8. (a) DCM performance for each DCM resource combinations (2020). (b) An exampl of the payback effect (Aug. 25, 2020).

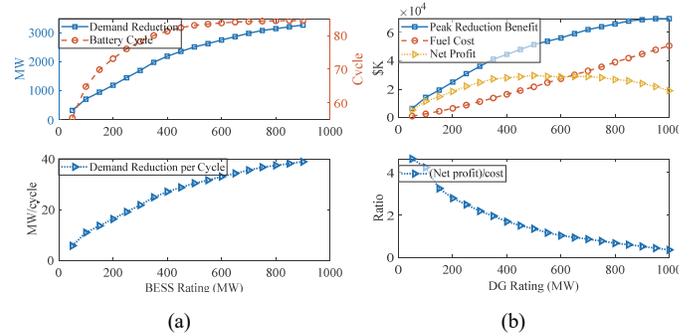

Fig. 9. (a) Sensitivity analysis for different BESS ratings and (b) Sensitivity analysis for different DG ratings.